# Single-beam all-optical non-zero field magnetometric sensor for magnetoencephalography applications

M.V. Petrenko[1], A.S. Pazgalev[1], and A.K. Vershovskii[1]

[1]*Ioffe Institute, Russian Academy of Sciences, St. Petersburg, 194021 Russia*
*e-mail address: antver@mail.ioffe.ru*

We present a method for measuring the magnetic field that allows hyperfine and Zeeman optical pumping, excitation and detection of magnetic resonance by means of a single laser beam with time-modulated ellipticity. This improvement allows us to significantly simplify the Bell-Bloom magnetometric scheme, while retaining its sensitivity. The method does not require the use of radio frequency fields, which is essential when creating arrays of sensors. The results of experimental studies demonstrate the efficiency of the proposed method and its potential applicability in most challenging magnetoencephalographic tasks.

**Keywords:** optically detectable magnetic resonance, parametric excitation of magnetic resonance, single-beam scheme, quantum magnetometer, all-optical magnetometer, non-zero magnetic field sensor

## INTRODUCTION

One of the most urgent problems of modern magnetometry in its application to the study of magnetic activity of the human brain is the creation of a compact non-cryogenic magnetic field sensor that can function in non-zero magnetic fields (MF) [1]. This requirement is due, to a large extent, to the drawbacks of magnetically shielded rooms capable of providing zero MF (hereinafter, by zero field we mean a field not exceeding one-two hundred nT in absolute value) inside a volume sufficient for magnetoencephalographic (MEG) studies: the extremely high cost of these rooms, the complexity of their installation and maintenance, and their complete immobility.

Optically pumped magnetic field sensors (often also called "atomic magnetometers") are used in a wide range of applications including fundamental physics, mineral exploration, and biomedical imaging. The highest sensitivity to date has been achieved using the spin-exchange relaxation free (SERF) mode [2–5], which can be implemented in zero MF. The compactness of SERF sensors, and primarily of their commercial version, QuSpin [6,7] (*https://quspin.com*), relative to existing optical nonzero-field sensors, is achieved primarily due to the fact that they are based on the Hanle effect [8,9], and are therefore able to use only one laser for both optical pumping (OP) of the atomic medium and for optical detecting (OD) of magnetic resonance (MR) in this medium. However, it should be noted that SERF sensors must use low frequency modulation of the measured MF, which can be an obstacle to their operation in the array.

Schemes operating in the regime of partial light suppression of spin-exchange broadening [10,11] achieve somewhat worse sensitivity values. These sensors operate in nonzero MF under strong pumping: as atoms are concentrated on the magnetic levels that do not interact with light, the number of their possible spin-exchange partners decreases (the so-called stretched state [12]), as does the rate of spin exchange. Such schemes have achieved sensitivity values that make them usable in MEG systems [11,13,14]. Sensors are being developed based on both coated [15,16] and buffer gas-filled [17,18] cells.

The classic and most sensitive version of the scheme of a nonzero field sensor based on alkali atoms in the gas phase requires the use of two lasers: the circularly polarized light of one laser is tuned to the optical transition and implements the OP of the atoms, while the linearly polarized light of the second laser is detuned from the optical transition and performs OD based on rotation of the plane of its polarization [19,20].

Simplified versions of this scheme are widespread: MR detection can be carried out by absorption of a circularly polarized resonant beam, while the OP and OD beams can be combined into a single beam parallel to MF (so-called $M_Z$ scheme [11,21]), or a beam directed at an angle to the MF (single-beam $M_X$ scheme [22]). In all cases, the simplification of the circuit is achieved at the expense of either speed or sensitivity. As a result, single-beam sensors of a nonzero field are of little use for MEG applications.

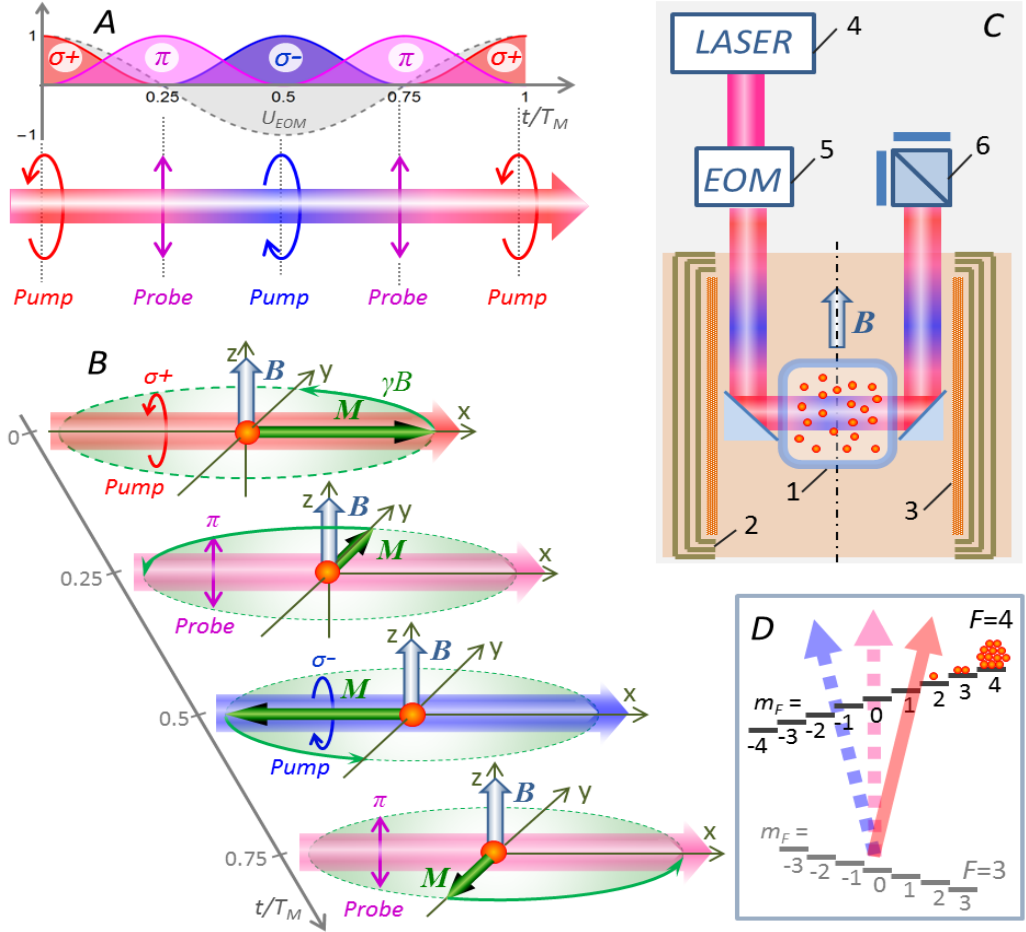

FIG. 1. (Color online) (A) Single-beam modulation period: σ± – circular polarizations, π – linear polarization. (B) The phases of the precession of the collective magnetic moment ***M***. At resonance, the precession of ***M*** is precisely synchronized with the optical pump modulation; at the moments when the beam is linearly polarized and is used as a probe, the projection of ***M*** onto the direction of the beam (x-axis) is zero. When the modulation frequency is detuned from magnetic resonance, a phase shift between ***M*** and OP appears; as a consequence, so does the projection $M_X$ of the moment. (C) Simplified setup diagram: 1 – cell with Cs vapor, 2 – multilayer magnetic shield, 3 – solenoid, 4 – external cavity diode laser, 5 – EOM, 6 – balanced photodetector. (D) Scheme of hyperfine ($F = 3,4$) and Zeeman ($m_F = -F..F$) levels of the $6^2S_{1/2}$ ground state of Cs in the presence of modulated optical pumping.

In turn, the two-beam scheme can be implemented in two versions. In the "classic" implementation, the OP beam is directed along the MF vector, and the MR excitation is carried out by a resonant radio field [23,24]. In the Bell-Bloom implementation, the OP beam is directed perpendicularly to the MF vector and modulated in amplitude, frequency, or polarization [25–27].

The Bell-Bloom scheme has two serious advantages over the "classic" one when used in MEG systems. First, such sensors do not use RF fields to excite an MR, and therefore do not interfere with neighboring sensors in the array. Second, the OP and OD beams in them can be directed almost parallel to each other. Therefore, these sensors have the potential to be more compact, have fewer "blind" zones [28,29], and can be more freely oriented in relation to the direction of the MF.

However, in a compact scheme, the problem of combining two beams and then separating them (so that the modulated OP beam does not hit the photodetector) seems to be nontrivial. At the time of writing, solutions that involve the use of precision interference optics and lasers tuned to different ($D_1$ and $D_2$) absorption lines of the atomic medium have been proposed [18,30].

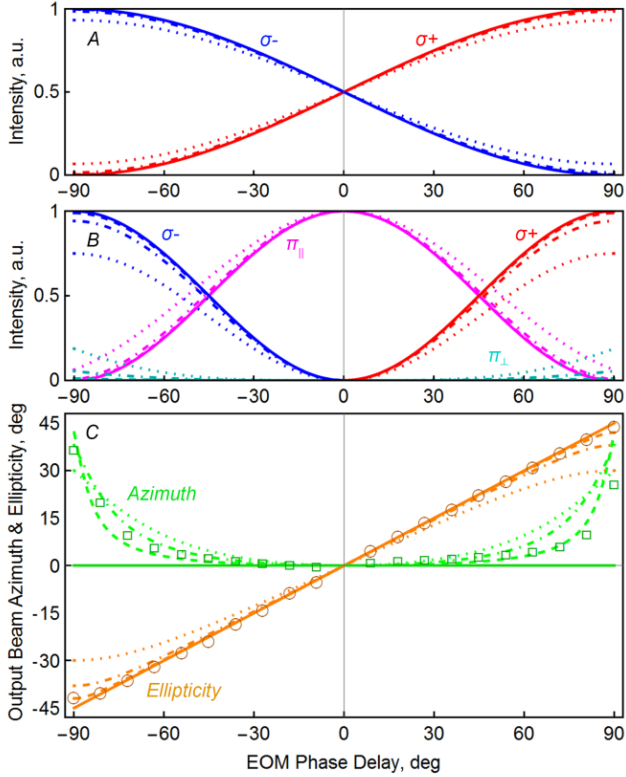

FIG. 2. (Color online) (A) Light at the EOM output, presented as the sum of two circular components, conventionally designated as σ– and σ+. (B) The same light, represented as a sum of two linear ($\pi_{\parallel}$ and $\pi_{\perp}$) and one alternating circular (σ±) components. (C) Azimuth and ellipticity (or phase delay) of the EOM output light. Solid lines correspond to a 45° angle of the EOM axis relative to the polarization azimuth of the incoming light, dashed lines – 42°, dash-dotted lines – 38°, dashed lines – 30°. Lines represent calculation, symbols – experiment.

The most notable among most recent studies on this subject is [15], in which a two-beam Bell-Bloom scheme with amplitude modulation of the pump beam in a coated cell was used to record brain activity in the Earth's field.

The best sensitivity for earth-scale field was demonstrated in [31] by means of detecting free spin precession (FSP) in a state-of-the-art multi-pass buffer gas cell. Two-beam detection method also was used in impressive experiment [32], demonstrating the detection of biomagnetic signals from the human brain in the ambient environment.

In contrast to all of the above, our work uses a *single* laser for optical pumping, magnetic resonance excitation, and detection. Magnetic resonance is excited by the modulation of circularly polarized component of the beam, and detection is achieved by the linearly polarized component in a quantum non-destructive manner. This improvement allows us to significantly simplify the Bell-Bloom scheme, while retaining all its sensitivity. The differences between our method and the methods used in previous works are further discussed below.

## DESCRIPTION OF THE METHOD

We propose modulating the laser beam so that it performs the functions of *i*) pumping (OP), *ii*) MR excitation, and *iii*) MR detection (OD). To do this, we propose varying the ellipticity $E$ of the beam in time from –1 to +1 (the corresponding phase delay angle between linear components takes values from –45° to +45°), i.e., from left-handed to right-handed circular polarization (Fig. 1A), with a frequency $\omega_M$ close to the Larmor frequency $\omega_0$. Note that this type of modulation can be easily achieved using an electro-optical modulator (EOM) when its axes are oriented at an angle of ±45° to the polarization plane of the incoming beam, and an oscillating (ramp or sinusoidal) voltage of the corresponding amplitude is applied to it.

In this setup, the beam's polarization becomes purely circular (σ+ or σ-) twice during the period $T_M=2\pi/\omega_M$; likewise, it becomes purely linear (π) twice during the period. Between these moments, the ellipticity takes intermediate values (Fig. 2). OP and the excitation of MR is carried out by the σ± component of the beam, while the detection is carried out by the π component; these two processes are decoupled in time, or, more precisely, in phase, within one modulation period.

A single OP/OD beam is tuned to a frequency close to the frequency of the optical transitions $F = I – ½ \leftrightarrow F' = I ± ½$ of the $S_{1/2}$ ground state of an alkali metal (in our experimental work, we use Cs, although the method is also applicable to other alkali metals, such as Rb and K). As shown in [10,33,34], and theoretically substantiated in [35], such a beam is capable of performing both Zeeman and hyperfine pumping. The beam depletes the $F = I –½$ level and reduces the optical density of the medium; in addition, it strongly polarizes the Zeeman structure of the $F = I +½$ level, forming the stretched state. As a result of hyperfine pumping, atoms are concentrated at the level $F = I + ½$, and therefore the π-component of the beam, which we use for OD (Fig. 1B), mainly detects the MR at a level from which it is detuned in frequency by an amount of the order of the hyperfine splitting of the ground state (for Cs it is 9.192 GHz). This provides the appropriate conditions for nondestructive quantum detection; thus, we achieve near-optimal conditions for both OP and OD.

It should be noted that the π-component of the beam also detects the MR at the level $F = I - \frac{1}{2}$. But, firstly, this level is almost depleted, and, secondly, the MR at this level is strongly broadened by resonant light, so its response can manifest itself only in the form of a wide pedestal.

In general, light wave with arbitrary polarization can be represented as the sum of either two linear or two circular (Fig. 2A) components, but in the framework of this paper, it is convenient to consider the light as a sum of a linear and alternating circular components (Fig. 2B). A simple calculation, in which the EOM is considered as a wave plate with a phase delay proportional to the applied voltage, shows that the polarization azimuth of the radiation transmitted through the EOM remains unchanged only when its axes are precisely oriented at an angle of ±45° to the polarization plane of the incoming beam. A deflection from ±45° leads to the appearance of a transverse (π⊥) component of linear polarization, or, in other words, to a rotation of the polarization of the π component when the control voltage is applied to the EOM. This rotation is detected by the photodetector as a "baseline" – a signal unrelated to MR, oscillating in-phase with modulation.

The evolution of the magnetic moment $M$ in the magnetic field $B$ is described by the Bloch equation

$$\frac{dM}{dt} = -\hat{\Gamma} \cdot M - |\gamma| M \times B + I_P(t) \cdot (M - M_P). \quad (1)$$

Here $\Gamma$ is the relaxation operator, $B$ is the magnetic field vector, $\gamma$ is the gyromagnetic ratio, $I_P(t) = I_+ - I_- = I_0 \cdot 2J_1(q) \cdot \sin(\omega_M t)$ is the time-dependent circular pumping rate, $M_P$ is the equilibrium value of the magnetic moment caused by the pumping, $I_0$ is the pump intensity, $J_1(q)$ is the Bessel function, $q$ is the phase modulation index at the output of the EOM.

By introducing rotating components $M_\pm = (M_X \pm iM_Y)$ and discarding rapidly oscillating terms, we obtain a stationary solution for the moment component $M_X$ collinear with the OP/OD beam direction:

$$M_X \sim \frac{I_0 \cdot 2J_1(q)}{(\Gamma_2 + I_0/2)^2 + (\omega_M - \omega_0)^2} \times \\ \times [\Gamma_2 \cdot \sin(\omega_M \cdot t) - (\omega_M - \omega_0) \cdot \cos(\omega_M \cdot t)]. \quad (2)$$

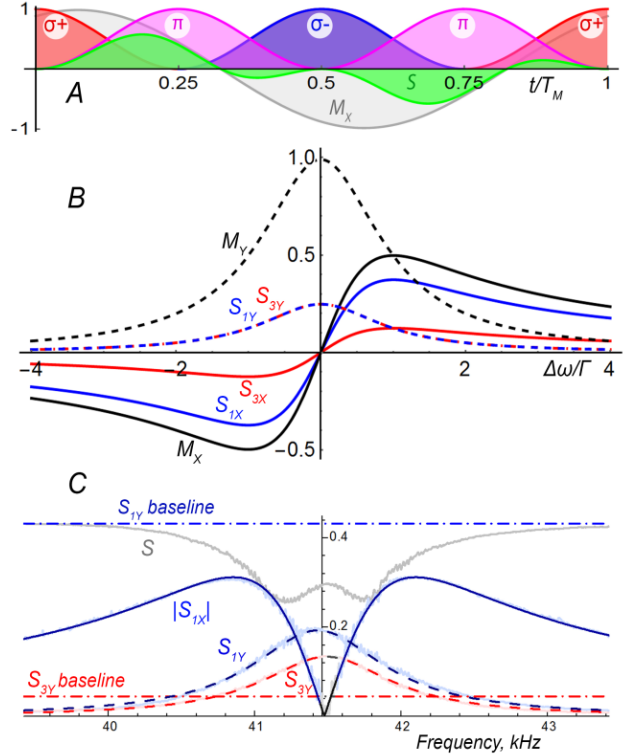

FIG. 3. (Color online) (A) One period of beam modulation: σ± and π polarizations, the projection of the magnetic moment $M$ onto the beam, and the signal $S$ at non-zero resonance detuning. (B) The $S_X$ and $S_Y$ components of the signal, obtained using non-modulated ($M_X$, $M_Y$) and modulated ($S_{1X}$, $S_{1Y}$, $S_{3X}$, $S_{3Y}$) OD beam (calculation on the basis of stationary solutions of the Bloch equations; the numbers in the indices correspond to the number of the signal harmonic). (C) Experimentally measured amplitudes of the 1st and 3rd harmonics of $S_X$ and $S_Y$ components of the signal; smooth lines are the result of approximation and serve as a guideline to the eye.

Note the difference between the Bell-Bloom scheme and the "classic" scheme that uses an RF excitation: at exact resonance, the collective magnetic moment $M$ in the latter precesses at an angle $\pi/2$ with respect to the magnetic component of the exciting field. In the version of the Bell-Bloom scheme presented here, the parametric resonance is excited in phase with the pump modulation. At resonance, the precession is precisely synchronized with the pumping (σ±) component of the beam; therefore, at moments when the amplitude of the detecting π-component of the beam is at a maximum, the projection $M_X$ is zero. When the modulation frequency $\omega_M$ is detuned from the resonance frequency $\omega_0$, a phase difference between **M** and OP appears, resulting in nonzero projection $M_X$.

Nonzero $M_X$ component, in turn, causes a rotation of the polarization plane of the π-component of the beam (Fig. 3A). The signal, proportional to the first harmonic of the polarization plane rotation, can be estimated by multiplying the intensity of π-component of the beam

$$I_\pi \sim J_0(q) - 2J_2(q) \cdot \cos(2 \cdot \omega_M \cdot t) \qquad (3)$$

by the value of $M_X$ component given by Eq.2. The magnitude of the rotation of the polarization plane is proportional to $[J_0(q) - J_2(q)] \cdot J_1(q)$ and maximal at the phase modulation index $q \sim 0.9$, corresponding to the EOM phase delay about 52°.

It follows from Eq.2,3 that since the intensity of the π-component of the beam is modulated at $2 \cdot \omega_M$, the 3rd harmonic appears in the signal in addition to the 1st harmonic. In case of linear phase modulation the *x*-components of these harmonics are equal, respectively, to ¾ and ¼ of the value of the *x*-component of the signal in the scheme using non-modulated OD beam (Fig. 3B).

As mentioned above, the rotation of the polarization azimuth of the π-component, which appears when the EOM is inaccurately tuned, leads to the appearance of "baselines" at the frequencies of the 1st (and, as a consequence, 3rd) harmonics of the modulation frequency in the signal (Fig. 3C). But, firstly, their magnitudes can be minimized, and, secondly, they oscillate in phase with the *y*-components of the signal, and their presence does not lead to a shift of the *x*-components which are of interest to us.

Now let us discuss the differences between the proposed method and the previously suggested approaches. The method proposed in [3] is the closest prototype of our method: a sensor using non-modulating, elliptical polarization for both OP and OD. The circular component of the light was used for spin polarization, while the linear component was used to measure optical rotation. However, this sensor is a SERF type, it only works in zero fields, and is not fully optical due to the field modulation necessary for detection.

Unlike [3], we pump resonance with maximum efficiency with circularly polarized light (twice per period of the Larmor frequency), and perform resonance detection in the intervals between pulses (also twice per period).

A modulation of circular polarization in single-beam scheme was also proposed in [28,36,37,26]. In [28] modulated ellipticity was used in a buffer gas cell but, in contrast to our approach, the laser was tuned to the level $F = I + ½$, which significantly limited the possibility of achieving a "stretched" state. For the first time, polarization push-pull pumping was suggested in [36,37]; the polarization modulation at hyperfine frequencies required using Mach–Zehnder interferometer. In [26], magnetic resonance in a cell with anti-relaxation coating was excited with rectangular modulation of circular polarization.

The best sensitivity to the Earth-scale field was demonstrated in [28] by recording the free spin precession (FSP) in a modern multi-pass cell with gas with two beams pump-probe measurement scheme.

All these studies [26,28,36,37] used the absorption signal for detection, so detection was not of quantum non-demolition type. Moreover, in schemes that use absorption detection together with modulation of the light parameters, it is very difficult to separate the useful signal from the initial beam modulation.

In our work, the signal is detected by rotating the plane of polarization of linearly polarized light. The use of the light tuned to the optical transition from $F = I - ½$ level in a single-beam system makes it possible to achieve non-demolition detection of the transition from $F = I + ½$ level, significantly suppressing the influence of the laser intensity noise [19,38].

In the most notable among most recent studies are [15,32], two-beam Bell-Bloom schemes were used to record brain activity in the Earth's field. In [15], a Cs paraffin-coated cell with a diameter of 25 mm, operating at room temperature, was used. The pump power was modulated by an acoustic-optical modulator.

In an impressive experiment [32], the FSP signal was detected by linear polarization rotation in two-beam scheme. A gradiometer consisting of two multi-pass Rb cells 8×8×12.5 mm in size has demonstrated the sensitivity level about 15 fT/√Hz/cm outdoors.

By contrast to [15,32], in our work, a single laser is used for OP, excitation of MR, and OD. Another significant difference from work [32] is our scheme's ability to operate both in the pulsed FSP mode and in the continuous optically-driven spin precession (ODSP) mode. Switching between FSP and ODSP modes does not require the introduction of additional optoelectronic devices into the scheme.

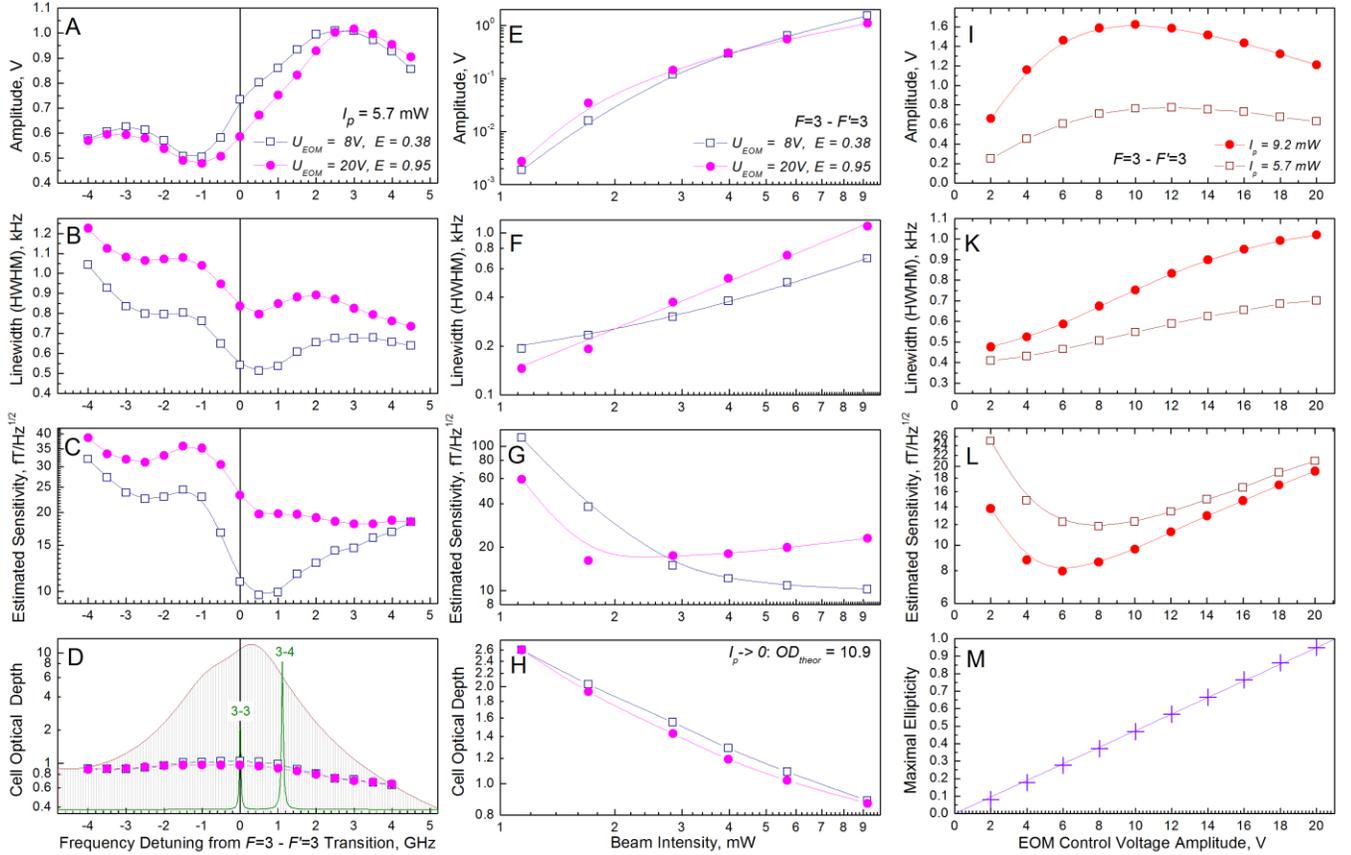

FIG. 4. (Color online) Experimentally measured dependences of the signal amplitude (A, E, I), resonance width (B, F, K), ultimate (limited by shot noise of the detecting light) sensitivity (C, G, L), optical cell thickness (D, H), and the maximal ellipticity of the beam (M) from the detuning of the beam frequency relative to the transition $F = 3 \leftrightarrow F' = 3$ of the Cs $D_1$ line (A – D), the beam intensity (E – H), and the amplitude of the EOM control voltage (further amplified ten times by a high-voltage amplifier) (I – M). Also shown are the calculated Cs absorption profiles in a vacuum cell and a cell filled with nitrogen at a pressure of 100 torr (D). The conditions under which the experiments were carried out are indicated in the upper cells of each column.

## EXPERIMENTAL

The experimental setup (Fig. 1C) is described in [34]. We use a cubic glass cell with internal dimensions of 8×8×8 mm manufactured by the VitaWave company, containing saturated Cs vapor at a temperature of about 95°C (corresponding to the number density of ~ $10^{13}$ cm$^{-3}$) and nitrogen under a pressure of ~100 torr. The cell was placed in a multi-layer magnetic shield, in which a MF induction of ~12 µT was maintained. An EOM (Thorlabs EO-AM-NR-C1) was used to modulate the ellipticity of the transverse OP/OD beam. The control voltage before being fed to the EOM was amplified ten times by a high-voltage amplifier Thorlabs HVA200. The signal was detected by a balanced photodetector; synchronous detection allowed us to separate the $x$-component of the signal, the amplitude of which is zero at the center of the resonance (as shown above).

The difference between our experiment and [34] is in the use of a sinusoidal (or symmetric triangular) control voltage on the EOM, and in the detection of resonance via the rotation of the polarization plane of the same beam that is used for pumping. At this stage, we investigated the parameters of the $M_X$ resonance signals in this scheme, and evaluated its ultimate sensitivity, limited by the shot noise level [39].

The results of the experiment are shown in Fig. 4. It can be seen that for all the main parameters there are optima that make it possible to achieve maximum sensitivity. In particular, the laser frequency optimum almost coincides with the maximum of the absorption line (Fig. 4C, D), since at this condition the atomic medium achieves the maximum degree of orientation, and, accordingly, the narrowing of the resonance due to suppression of spin-exchange broadening [9] (Fig. 4B), as well as bleaching of the medium (Fig. 4D). In this case, the detuning of laser radiation from the transitions $F = 4 \leftrightarrow F' = 3,4$, which it interrogates, is approximately 10 GHz. The optimum in terms of the ellipticity of the beam was realized at a value of 0.3–0.4, with a significant excess of the intensity of the detecting component over the pumping one (Fig. 4L). The best value of the expected ultimate sensitivity was found to be below 8 fT/√Hz. Simultaneous detection of the 1$^{st}$ and 3$^{rd}$ harmonics of the signal will improve this the value by one third (Fig.3B). As follows from Fig.2A,B, a sinusoidal change in the intensities of the beam polarization components is achieved with a linear change in the control voltage on the EOM; therefore, the transition from a sinusoid modulation to the linear (ramp) one makes it possible to increase the fraction of the 1$^{st}$ harmonic of the modulation, and, as the experiment has shown, to increase the signal amplitude by about 20% more with about 5% lesser resonance width.

Of particular note is the smaller width of resonances realized in the proposed single-beam scheme compared with the widths obtained earlier in the same cell in two-beam schemes [13], which is explained by the absence of resonance broadening by the second (detecting) beam and the radio-frequency field.

To independently verify our understanding of the principles of signal shaping in this scheme, we also applied square-wave (meander) modulation to the EOM. As expected, the absence of a detecting π component in the beam led to zeroing of amplitude of the first harmonic of the MR signal.

Up to this point, we have considered the scheme for continuous excitation of magnetic resonance; however, the pumping and detection scheme proposed in this work can also be used to excite and register free spin precession signals. Moreover, technically, the transition to FSP is carried out by simply turning off the control signal on the EOM; in this case, the FSP signal is recorded by a non-modulated beam, and therefore contains only the first harmonic. FSP schemes [32] are characterized by both disadvantages (due to the pulsed nature of their operation), and advantages (for example, the absence of a MR frequency shifts by the pump light and errors associated with the phase alignment in the feedback loop [40]).

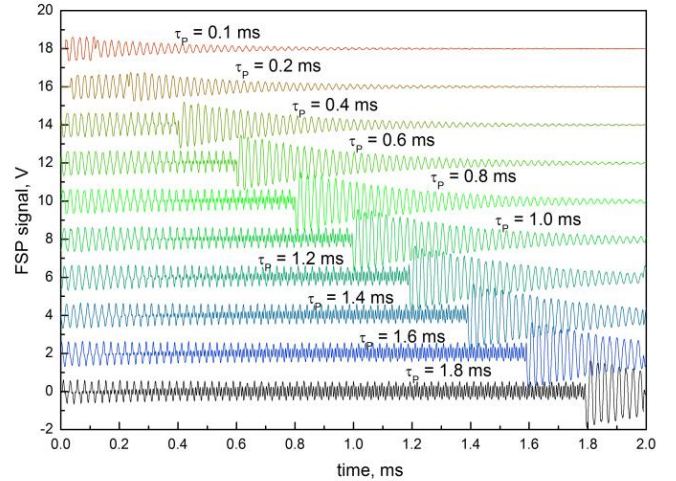

FIG. 5. (Color online) Time dependences of free precession signals at different optical pumping intervals.

In the proposed version, the transition to the FSP registration is carried out with maximum simplicity without introducing additional control elements into the circuit. We also note the absence of harmonics and baselines in the FSP signal: they disappear when the control voltage on the EOM is turned off.

## CONCLUSIONS

We have experimentally demonstrated that optical pumping, excitation and detection of magnetic resonance can be carried out very efficiently using a single laser beam with time-modulated ellipticity in a nonzero field magnetometric sensor. Our research confirms that the method is not only efficient, but also capable of providing metrological parameters that are comparable to the best existing schemes developed for nonzero field magnetoencephalographic systems. The efficiency of the scheme in the pulsed mode of excitation and registration of signals of free spin precession was also demonstrated. Certain technical problems can arise when transmitting a beam with modulated ellipticity over an optical fiber; as a last resort, the EOM can be located at the fiber's exit. But even so, the exclusion of the second laser from the scheme provides much greater simplicity and, as a consequence, compactness of the sensor.

## ACKNOWLEDGEMENTS

The reported study was funded by RFBR, project number 19-29-10004.

# REFERENCES

[1] K.-M. C. Fu, G. Z. Iwata, A. Wickenbrock, and D. Budker, *Sensitive Magnetometry in Challenging Environments*, AVS Quantum Sci. **2**, 044702 (2020).

[2] J. C. Allred, R. N. Lyman, T. W. Kornack, and M. V. Romalis, *High-Sensitivity Atomic Magnetometer Unaffected by Spin-Exchange Relaxation*, Phys. Rev. Lett. **89**, 130801 (2002).

[3] V. Shah and M. V. Romalis, *Spin-Exchange Relaxation-Free Magnetometry Using Elliptically Polarized Light*, Phys. Rev. A **80**, 1 (2009).

[4] M. P. Ledbetter, I. M. Savukov, V. M. Acosta, D. Budker, and M. V. Romalis, *Spin-Exchange-Relaxation-Free Magnetometry with Cs Vapor*, Phys. Rev. A **77**, 033408 (2008).

[5] H. B. Dang, A. C. Maloof, and M. V. Romalis, *Ultrahigh Sensitivity Magnetic Field and Magnetization Measurements with an Atomic Magnetometer*, Appl. Phys. Lett. **97**, 151110 (2010).

[6] T. M. Tierney, N. Holmes, S. Mellor, J. D. López, G. Roberts, R. M. Hill, E. Boto, J. Leggett, V. Shah, M. J. Brookes, R. Bowtell, and G. R. Barnes, *Optically Pumped Magnetometers: From Quantum Origins to Multi-Channel Magnetoencephalography*, NeuroImage **199**, 598 (2019).

[7] E. Boto, S. S. Meyer, V. Shah, O. Alem, S. Knappe, P. Kruger, T. M. Fromhold, M. Lim, P. M. Glover, P. G. Morris, R. Bowtell, G. R. Barnes, and M. J. Brookes, *A New Generation of Magnetoencephalography: Room Temperature Measurements Using Optically-Pumped Magnetometers*, NeuroImage **149**, 404 (2017).

[8] A. Kastler, *The Hanle Effect and Its Use for the Measurements of Very Small Magnetic Fields*, Nuclear Instruments and Methods **110**, 259 (1973).

[9] G. Le Gal, G. Lieb, F. Beato, T. Jager, H. Gilles, and A. Palacios-Laloy, *Dual-Axis Hanle Magnetometer Based on Atomic Alignment with a Single Optical Access*, Phys. Rev. Applied **12**, 064010 (2019).

[10] T. Scholtes, V. Schultze, R. IJsselsteijn, S. Woetzel, and H.-G. Meyer, *Light-Narrowed Optically Pumped ${M}_{x}$ Magnetometer with a Miniaturized Cs Cell*, Phys. Rev. A **84**, 043416 (2011).

[11] V. Schultze, B. Schillig, R. IJsselsteijn, T. Scholtes, S. Woetzel, and R. Stolz, *An Optically Pumped Magnetometer Working in the Light-Shift Dispersed Mz Mode*, Sensors **17**, 3 (2017).

[12] S. Appelt, A. Ben-Amar Baranga, A. R. Young, and W. Happer, *Light Narrowing of Rubidium Magnetic-Resonance Lines in High-Pressure Optical-Pumping Cells*, Phys. Rev. A **59**, 2078 (1999).

[13] A. K. Vershovskii, A. S. Pazgalev, and M. V. Petrenko, *All-Optical Magnetometric Sensor for Magnetoencephalography and Ultralow Field Tomography*, Tech. Phys. Lett. **46**, 877 (2020).

[14] A. E. Ossadtchi, N. K. Kulachenkov, D. S. Chuchelov, S. P. Dmitriev, A. S. Pazgalev, M. V. Petrenko, and A. K. Vershovskii, *Towards Magnetoencephalography Based on Ultrasensitive Laser Pumped Non-Zero Field Magnetic Sensor*, in *2018 International Conference Laser Optics (ICLO)* (2018), pp. 543–543.

[15] R. Zhang, W. Xiao, Y. Ding, Y. Feng, X. Peng, L. Shen, C. Sun, T. Wu, Y. Wu, Y. Yang, Z. Zheng, X. Zhang, J. Chen, and H. Guo, *Recording Brain Activities in Unshielded Earth's Field with Optically Pumped Atomic Magnetometers*, Science Advances **6**, eaba8792 (2020).

[16] H. Wang, T. Wu, W. Xiao, H. Wang, X. Peng, and H. Guo, *Dual-Mode Dead-Zone-Free Double-Resonance Alignment-Based Magnetometer*, Phys. Rev. Applied **15**, 024033 (2021).

[17] V. G. Lucivero, W. Lee, N. Dural, and M. V. Romalis, *Femtotesla Direct Magnetic Gradiometer Using a Single Multipass Cell*, Phys. Rev. Applied **15**, 014004 (2021).

[18] A. R. Perry, M. D. Bulatowicz, M. Larsen, T. G. Walker, and R. Wyllie, *All-Optical Intrinsic Atomic Gradiometer with Sub-20 FT/Cm/√Hz Sensitivity in a 22 µT Earth-Scale Magnetic Field*, Opt. Express, OE **28**, 36696 (2020).

[19] D. Budker, W. Gawlik, D. F. Kimball, S. M. Rochester, V. V. Yashchuk, and A. Weis, *Resonant Nonlinear Magneto-Optical Effects in Atoms*, Rev. Mod. Phys. **74**, 1153 (2002).

[20] W. Gawlik and S. Pustelny, *Nonlinear Magneto-Optical Rotation Magnetometers*, in *High Sensitivity Magnetometers*, edited by A. Grosz, M. J. Haji-Sheikh, and S. C. Mukhopadhyay (Springer International Publishing, Cham, 2017), pp. 425–450.

[21] A. L. Bloom, *Principles of Operation of the Rubidium Vapor Magnetometer*, Appl. Opt., AO **1**, 1 (1962).

[22] S. Groeger, G. Bison, J.-L. Schenker, R. Wynands, and A. Weis, *A High-Sensitivity Laser-Pumped Mx Magnetometer*, Eur. Phys. J. D **38**, 239 (2006).

[23] H. G. Dehmelt, *Modulation of a Light Beam by Precessing Absorbing Atoms*, Phys. Rev. **105**, 1924 (1957).

[24] W. Chalupczak, R. M. Godun, P. Anielski, A. Wojciechowski, S. Pustelny, and W. Gawlik, *Enhancement of Optically Pumped Spin Orientation via Spin-Exchange Collisions at Low Vapor Density*, Phys. Rev. A **85**, 043402 (2012).

[25] R. Gartman and W. Chalupczak, *Amplitude-Modulated Indirect Pumping of Spin Orientation in*

*Low-Density Cesium Vapor*, Phys. Rev. A **91**, 053419 (2015).

[26] Z. D. Grujić and A. Weis, *Atomic Magnetic Resonance Induced by Amplitude-, Frequency-, or Polarization-Modulated Light*, Phys. Rev. A **88**, 012508 (2013).

[27] Y. Guo, S. Wan, X. Sun, and J. Qin, *Compact, High-Sensitivity Atomic Magnetometer Utilizing the Light-Narrowing Effect and in-Phase Excitation*, Appl. Opt., AO **58**, 4 (2019).

[28] A. Ben-Kish and M. V. Romalis, *Dead-Zone-Free Atomic Magnetometry with Simultaneous Excitation of Orientation and Alignment Resonances*, Phys. Rev. Lett. **105**, 193601 (2010).

[29] G. Bao, A. Wickenbrock, S. Rochester, W. Zhang, and D. Budker, *Suppression of the Nonlinear Zeeman Effect and Heading Error in Earth-Field-Range Alkali-Vapor Magnetometers*, Phys. Rev. Lett. **120**, 3 (2018).

[30] V. Gerginov, M. Pomponio, and S. Knappe, *Scalar Magnetometry Below 100 FT/Hz1/2 in a Microfabricated Cell*, IEEE Sensors Journal **20**, 12684 (2020).

[31] D. Sheng, S. Li, N. Dural, and M. V. Romalis, *Subfemtotesla Scalar Atomic Magnetometry Using Multipass Cells*, Phys. Rev. Lett. **110**, 160802 (2013).

[32] M. E. Limes, E. L. Foley, T. W. Kornack, S. Caliga, S. McBride, A. Braun, W. Lee, V. G. Lucivero, and M. V. Romalis, *Portable Magnetometry for Detection of Biomagnetism in Ambient Environments*, Phys. Rev. Applied **14**, 011002 (2020).

[33] W. Chalupczak, R. M. Godun, P. Anielski, A. Wojciechowski, S. Pustelny, and W. Gawlik, *Enhancement of Optically Pumped Spin Orientation via Spin-Exchange Collisions at Low Vapor Density*, Phys. Rev. A **85**, 043402 (2012).

[34] A. K. Vershovskii, S. P. Dmitriev, G. G. Kozlov, A. S. Pazgalev, and M. V. Petrenko, *Projection Spin Noise in Optical Quantum Sensors Based on Thermal Atoms*, Tech. Phys. **65**, 1193 (2020).

[35] E. N. Popov, V. A. Bobrikova, S. P. Voskoboinikov, K. A. Barantsev, S. M. Ustinov, A. N. Litvinov, A. K. Vershovskii, S. P. Dmitriev, V. A. Kartoshkin, A. S. Pazgalev, and M. V. Petrenko, *Features of the Formation of the Spin Polarization of an Alkali Metal at the Resolution of Hyperfine Sublevels in the 2S1/2 State*, Jetp Lett. **108**, 513 (2018).

[36] Y.-Y. Jau, E. Miron, A. B. Post, N. N. Kuzma, and W. Happer, *Push-Pull Optical Pumping of Pure Superposition States*, Phys. Rev. Lett. **93**, 160802 (2004).

[37] A. B. Post, Y.-Y. Jau, N. N. Kuzma, and W. Happer, *Amplitude- versus Frequency-Modulated Pumping Light for Coherent Population Trapping Resonances at High Buffer-Gas Pressure*, Phys. Rev. A **72**, 033417 (2005).

[38] V. Shah, G. Vasilakis, and M. V. Romalis, *High Bandwidth Atomic Magnetometery with Continuous Quantum Nondemolition Measurements*, Phys. Rev. Lett. **104**, 013601 (2010).

[39] D. Budker and M. Romalis, *Optical Magnetometry*, Nature Physics **3**, 227 (2007).

[40] A. K. Vershovskiĭ and E. B. Aleksandrov, *Phase Error Elimination in the Mxmagnetometer and Resonance Line Shape Control in an Unstable Field Using the Technique of Invariant Mapping of a Spin Precession Signal*, Opt. Spectrosc. **100**, 12 (2006).